# A Probabilistic Framework for Power System Large-Disturbance Global Instability Risk Assessment in the Presence of Renewable Wind Generation

**Umair Shahzad**

Department of Electrical and Computer Engineering,

University of Nebraska-Lincoln, Lincoln, NE, USA.

Email: umair.shahzad@huskers.unl.edu

**Abstract**

The increasing demand of large scale wind integration in the conventional power system brings a lot of challenges. One of them is the stability of the power system when subjected to a large disturbance, such as a fault. This paper proposes a probabilistic risk-based framework for computing a global instability index, incorporating angle, voltage, and frequency stability, for a large disturbance. Moreover, the impact of high wind penetration on this index is also observed. Case studies and associated simulations are conducted on the IEEE 39-bus test system using DIgSILENT PowerFactory software. The results show that higher penetration of wind generation enhances the global stability of the power system. Moreover, the impact of changing system generation and load is studied on the global instability index.

## 1. Introduction

Since the early 20$^{th}$ century, power system stability has been documented as a noteworthy subject in securing power system planning and operation [1-2]. Most of the blackouts caused by power system instability have demonstrated the significance of this phenomenon [3-4]. Historically, transient stability has been the leading stability issue in most power networks. However, with the introduction of novel technologies and increasing load demands, several kinds of instability have arisen. For instance, voltage stability, frequency stability and interarea oscillations have gained importance. This has necessitated an understanding of the basics of power system stability. A lucid concept of various kinds of instability is vital for the acceptable operation of power systems. [5] has broadly classified power system stability into three major kinds: frequency, voltage, and rotor angle. This is pictorially shown in Fig. 1. A brief description of these classifications follows.

*Frequency stability* is the ability of a power system to maintain steady frequency after a severe system stress causes a substantial disparity between generation and load. It relies on the ability to maintain equilibrium between system generation and demand, with minimum inadvertent loss of load. Instability can manifest itself in the shape of sustained frequency swings, which subsequently cause tripping of generating units and loads.

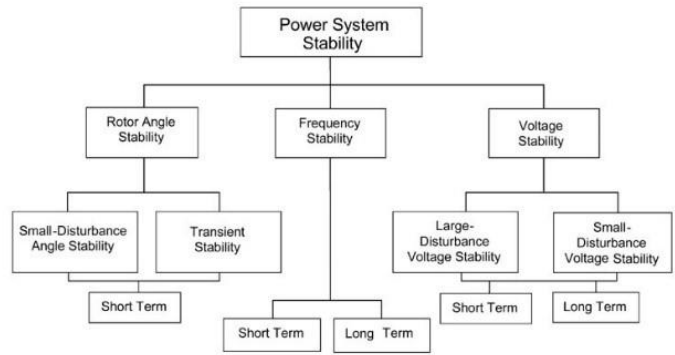

Figure 1. Classification of power system stability

*Voltage stability* is the ability of a power system to maintain steady voltages at all buses in the system, after being subjected to a disturbance from a given initial operating condition. It depends on the ability to maintain equilibrium between system demand and system generation. Instability may manifest itself in the shape of a continuing decrease or increase of voltages at some or all buses. The culprit for voltage instability is typically the loads. Large-disturbance voltage stability is the ability of a power system to maintain steady voltages after the occurrence of large disturbances, such as three-phase faults. The inherent features of system and load are major determinants of this ability. The period of interest typically ranges from a few seconds to tens of minutes. Small-disturbance voltage stability is the system's ability to maintain steady voltages when small disturbances, such as steady changes in network load, take place.

*Rotor angle stability* is the ability of synchronous machines in a power system to maintain synchronism when a disturbance is applied. Instability can result when the angular swing of generators causes a loss in synchronism. Small-signal rotor angle stability deals with stability under small disturbances, such as minor load variations. Large-disturbance rotor angle stability, or much more commonly called transient stability, focuses on the ability of the power system to maintain synchronism when a severe disturbance, such as a three-phase short circuit on a transmission line, is applied. The resulting system response constitutes large excursions of generator rotor angles and is affected by the nonlinear power-angle relation. Transient stability relies, for the most part, on the initial operating state of the system and the severity of the disturbance. Instability is normally manifested in the form of aperiodic angular separations which are due to inadequate synchronizing torque. The time range for transient stability studies is about 3 to 5 seconds after the disturbance has occurred. It may range up to 10-20 seconds for much larger systems with leading inter-area swings.

In conventional power system stability analysis, deterministic stability limit is used which depends on the worst-case scenario. This confines the feasible stable operating condition and thus, restricts the technical capabilities of the power generation to meet the load demand [6]. Moreover, the deterministic approach for power system stability assessment does not give any information on the status of present operating point but only provides information regarding the stability or instability of a current condition of operation [7]. Moreover, with the increasing usage of renewables such as wind and photovoltaics (PV), in addition to load uncertainties, assessing the stability in a deterministic manner is no longer pertinent [8].

## 2. Literature Review and Background

[9] elaborates some characteristics of the Tunisian power system transient stability. Angular stability is analyzed using the fault clearing time (FCT). Many contingency cases are considered: the outage of the largest generator and loss of load, along with insufficient operation of power system stabilizers (PSS). Although, the work considers different FCTs but only a three-phase fault at a specific line is chosen. Moreover, the integration of any renewable energy source is not part of this work. [10] proposes a probabilistic framework for power system transient stability assessment (TSA) with high renewable generation penetration. The presented framework enables comprehensive calculation of transient stability of power systems with abridged inertia. A major drawback of the work is the ignorance of bus faults and only considering three-phase line faults. Moreover, impact of different wind technologies and penetration levels are not considered. [11] presents a study of the effects of some important power system parameters on transient stability. The parameters considered for this assessment include fault location, load increment, machine

damping factor, FCT and generator synchronous speed. The work only considers three-phase line faults and does not study the influence of renewable generation integration on transient stability.

The transient stability analysis including wind power intermittency and volatility is proposed in [12]. The kinds of wind turbines considered in this work include the doubly fed induction generator (DFIG) and direct-driven permanent magnet synchronous generator (PMSG). The Monte Carlo (MC) simulation technique, joint with two evaluation indices, is tested on the IEEE 39-bus test system and a real China-Jiangxi power grid to investigate the impacts of intermittency and volatility of wind. Although, the work incorporates the effect of wake effect on transient stability, but it only considers a three- phase fault at a specific bus, which is cleared after a pre-selected time. [13] uses risk-based approach to analyze the transient stability of power networks incorporating wind farms. The proposed methodology of transient stability risk assessment is based on the MC method and eventually, an inclusive risk indicator, based on angle and voltage stability, has been devised. The work only considers three phase line faults and does not study the impact of different wind technologies. In [14], the abridged version of an altered single machine infinite bus (SMIB) system, with a DFIG-based wind farm integration, is analyzed using the transient features of the DFIG-based wind farm in the diverse periods of a fault. The assessment specifies that the working of synchronous generator can be either enhanced or depreciated with DFIG integration. Only a three-phase fault is considered on a specific line which clears after a pre-selected time.

In [15], the transient and voltage stability with high penetration levels of the diverse distributed generations (DGs) are examined. With each penetration level of DGs, the performance of power system is analyzed, and results are equated with the performance of power system base case (no DG integration). Only specific fault types and fault locations are considered. [16] explores the voltage stability of wind power grid integration. Furthermore, static synchronous compensator (STATCOM) and its sizing approach are proposed to improve the voltage stability of wind farm integrated grid. The results are valid for a specific fault and FCT. An approach for examining short-term voltage stability of power system incorporating induction motors is proposed in [17]. This approach considers one generator as the reference machine and the relative angle difference is considered as a differential variable. Besides, a mathematical approach of differential equation series is devised in the polar coordinates.

[18] studies the impact on transient and frequency stability for high wind penetration. DFIG wind generation is used and a sensitivity analysis is conducted for various wind generator loading conditions. The study proves that transient stability evaluation is highly dependent on fault location in the system. Frequency stability analysis shows that areas with reduced inertia are most impacted by generator outage events. [19] studies the impact of a large PV plant on the frequency stability of a power network when subjected to small and large disturbances. With the help of automatic generation control (AGC) and phasor measurement unit (PMU) data, the power generation levels of conventional generators are attuned to alleviate the deviations in system frequency. Additionally, the impact of increased PV penetration is inspected. Only a three-phase fault with a fixed FCT is considered.

In risk based stability assessment, the risk index is computed quantitatively for each possible contingency and its associated impact (severity) [20]. Some research work has been done in this area. For instance, [13] proposes a methodology to compute transient instability risk index based on angle and voltage instabilities. Although, the work considers the probabilistic nature of the FCT but the main drawback is the consideration of pre-determined fault locations of lines and associated probabilities. [21] proposes a methodology to compute the risk of transient instability of an operating point in a power network. The work includes various cost parameters to quantify the consequences of the risk.

From the literature review, some significant findings are obtained. Most works on probabilistic large-signal short term stability deals with only three phase faults. Although, the assumption may be suitable since it is the most severe fault, other faults cannot be ignored as their probabilities are higher, and must be included in probabilistic stability assessment. [22-23] indicate the need to establish a comprehensive approach for power system stability incorporating all stabilities: angle, voltage and frequency. [24-26] indicate that risk-based stability approach is an open area of research and requires further work. Moreover, the impact of reduced inertia systems (i.e. higher wind penetration) on power system stability is of great implication [27, 28]. Thus, the key contribution of this paper is to present a probabilistic approach for power system large-disturbance global instability risk assessment, in the presence of renewable wind generation, incorporating all kinds of stability (angle, voltage and frequency).

The rest of the paper is organized as follows. Section 3 discusses the mathematical formulation of risk-based global instability index for a large disturbance. Section 4 elaborates the procedure for computing this index. Section 5 briefly discussed the modelling approach for synchronous generators and DFIGs. Section 6 discuss case studies and associated simulations on the IEEE 39-bus test system. Section 7 presents the results and associated discussion. Finally, Section 8 concludes the paper with a suggested future research direction.

### 3. Mathematical Formulation of Risk-based Global Instability Index

This section will discuss the formulation of risk-based global instability index. The product of probability of an unforeseen event and its impact is commonly known as risk, which is mathematically defined as (1) [29, 30]. The reader can refer to [31] and [32] for detailed theory on risk and its applications in power systems.

$$Risk(X_{t,f}) = \sum_i \Pr(E_i)(\sum_j \Pr(X_{t,j}|X_{t,f}) \times Sev(E_i, X_{t,j})) \tag{1}$$

where $X_{t,f}$ is the forecasted operating condition at time $t$; $X_{t,j}$ is the $j^{th}$ possible loading level; $\Pr(X_{t,j}|X_{t,f})$ is the probability of $X_{t,j}$ given $X_{t,f}$ has occurred, which is obtained using a probability distribution for the possible load levels; $E_i$ is the $i^{th}$ contingency and $\Pr(E_i)$ is its probability; and $Sev(E_i, X_{t,j})$ quantifies the impact of $E_i$ occurring under the $j^{th}$ possible operating condition.

Let $R_A$ be the risk index for angle instability, i.e.

$$R_A(X_{t,f}) = \sum_{i=1}^{N} \Pr(F_i)(\sum_j P(X_{t,j}|X_{t,f}) \times Sev(A_i, X_{t,j})) \tag{2}$$

Where $\Pr(F_i)$ is given by

$$\Pr(F_i) = \Pr(F_{oi}) \times \Pr(F_{Li}) \times \Pr(F_{Ti}) \tag{3}$$

$\Pr(F_{oi})$, $\Pr(F_{Li})$, and $\Pr(F_{Ti})$ denote the probability of fault occurrence, fault location, and fault type, respectively for the $i^{th}$ MC sample. It is assumed that fault can occur on any line on the system and at any point along the line, hence,

$$\Pr(F_{oi}) = \frac{1}{N_L} \qquad (4)$$

where $N_L$ denotes total number of lines in the power system.

$$\Pr(F_{Li}) = \frac{1}{100} \qquad (5)$$

$\Pr(F_{Ti})$ denotes probability of fault type and will be chosen based on discrete distribution as described in Section 4.

$X_{t,f}$ is the forecasted operating condition at time $t$; $X_{t,j}$ is the $j^{th}$ possible loading level; $\Pr(X_{t,j} | X_{t,f})$ is the probability of this condition; $F_i$ is the $i^{th}$ fault and $\Pr(F_i)$ is its probability; and $Sev(A_i, X_{t,j})$ quantifies the impact of $F_i$ occurring under the $j^{th}$ possible operating condition. It is assumed that $Sev(A_i, X_{t,j})$ denotes the severity under peak loading condition. For simplicity, it is represented by $Sev(A_i)$ in the following equation.

$$Sev(A_i) = \begin{cases} |TSI_i|, & \text{if } TSI_i < 0 \\ 0, & \text{if } TSI_i > 0 \end{cases} \qquad (6)$$

where $TSI_i$ is the angle stability index i.e.

$$TSI_i = \frac{360 - \delta_{max\ i}}{360 + \delta_{max\ i}}, \quad -1 < TSI_i < 1 \qquad (7)$$

Where $\delta_{max\ i}$ is the maximum angle difference between any two synchronous generators after the fault at $i^{th}$ line. A negative $TSI_i$ indicates an unstable system. As an illustration, $\delta_{max}$ is indicated on Fig. 2 for a three-phase short circuit fault at Line 16-17 (line between bus 16 and bus 17) for the IEEE 39-bus system. The fault occurs at time $t=1.0$ s and clears at time $t=1.1$ s.

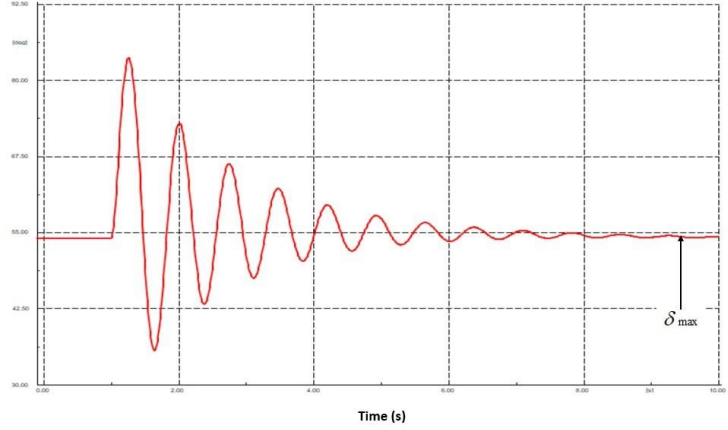

Figure 2.  Illustration of maximum rotor angle difference $\delta_{max}$

Let $R_V$ be the risk index for voltage instability, i.e.

$$R_V(X_{t,f}) = \sum_{i=1}^{N}\sum_{k=1}^{N_b} \Pr(F_i)(\sum_j P(X_{t,j} | X_{t,f}) \times Sev(V_i, X_{t,j})) \qquad (8)$$

$Sev(V_i, X_{t,j})$ quantifies the impact of $F_i$ occurring under the $j^{th}$ possible operating condition. It is assumed that $Sev(V_i, X_{t,j})$ denotes the severity under peak loading condition. For simplicity, it is represented by $Sev(V_i)$ in the following equation.

$$Sev(V_i) = \begin{cases} |V_{dk}|, & \text{if } |V_{dk}| > 5\% \\ 0, & \text{if } |V_{dk}| < 5\% \end{cases} \quad \forall k \in N_b \qquad (9)$$

$V_{dk}$ indicates the voltage deviation for bus $k$, i.e.

$$V_{dk} = V_{rated} - V_k \qquad (10)$$

Where $V_{rated}$ is the nominal voltage (1 pu) of each bus; $V_k$ is the post-fault steady-state voltage at $k^{th}$ bus; $N_b$ denotes total number of buses in the system, $N$ denotes number of MC samples.

As an illustration, $V_k$ (for Bus 20 of IEEE 39-bus system) is marked in Fig. 3 for a three-phase short circuit fault at Line 16-17. The fault occurs at time $t=1.0$ s and clears at time $t=1.1$ s.

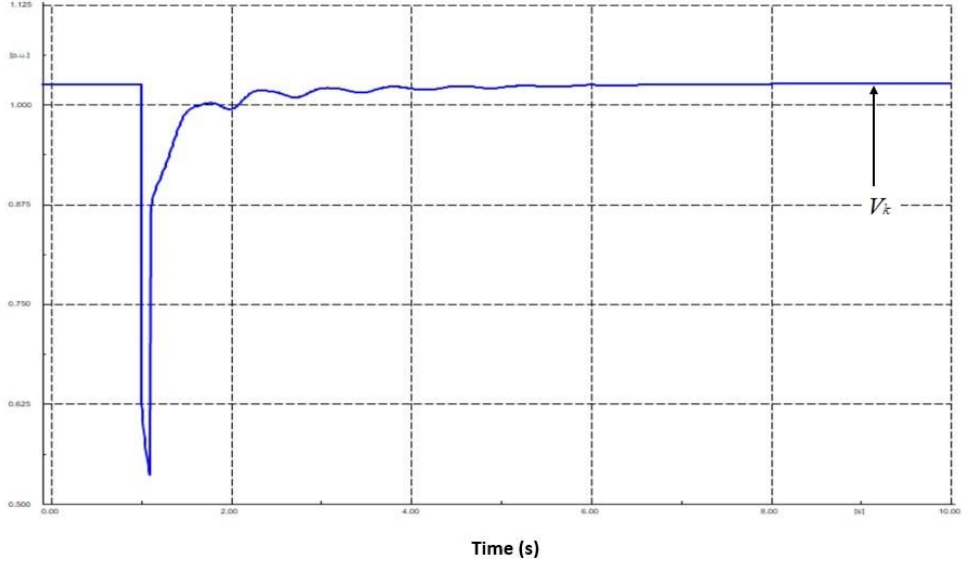

Figure 3. Illustration of post-fault steady state voltage $V_k$

Let $R_F$ be the risk index for frequency instability, i.e.

$$R_F(X_{t,f}) = \sum_{i=1}^{N}\sum_{k=1}^{N_g} \Pr(F_i)(\sum_{j} P(X_{t,j} \mid X_{t,f}) \times Sev(F_i, X_{t,j})) \qquad (11)$$

$Sev(F_i, X_{t,j})$ quantifies the impact of $F_i$ occurring under the $j^{th}$ possible operating condition. It is assumed that $Sev(F_i, X_{t,j})$ denotes the severity under peak loading condition. For simplicity, it is represented by $Sev(F_i)$ in the following equation.

$$Sev(F_i) = \begin{cases} |f_{dGk}|, & \text{if } |f_{dGk}| > 0.5Hz \\ 0, & \text{if } |f_{dGk}| < 0.5Hz \end{cases} \quad \forall G_k \in N_g \qquad (12)$$

$f_{dGk}$ denotes the post-fault deviation of frequency from nominal frequency (60 Hz) for $k^{th}$ generator; $N_g$ denotes total number of generators in the system. To illustrate, $f_{dGj}$ is marked in Fig. 4 for a three-phase short circuit fault at Line 16-17 (IEEE 39-bus system). The fault occurs at time $t=1.0$ s and clears at time $t=1.1$ s.

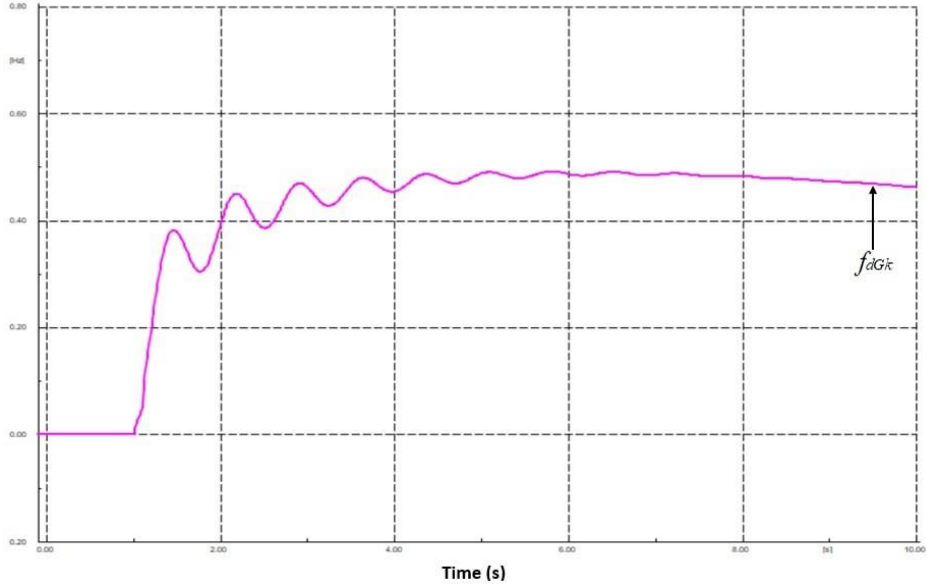

Figure 4. Illustration of post-fault electrical frequency deviation $f_{dGk}$

Let $R_{AM}$, $R_{VM}$ and $R_{FM}$ be the mean (average) values of $R_A$, $R_V$ and $R_F$, respectively. Mathematically,

$$R_{AM} = \frac{\sum_{n=1}^{N} R_{An}}{N} \quad (13)$$

$$R_{VM} = \frac{\sum_{n=1}^{N} R_{Vn}}{N} \quad (14)$$

$$R_{FM} = \frac{\sum_{n=1}^{N} R_{Fn}}{N} \quad (15)$$

Where $R_{An}$, $R_{Vn}$, $R_{Fn}$ are the values of $R_A$, $R_V$ and $R_F$ for $n^{th}$ MC sample.

Let $G$ be the global instability risk index, i.e.,

$$G = \max\{R_{AM}, R_{VM}, R_{FM}\} \quad (16)$$

## 4. Computation Procedure

There are various factors which are involved in probabilistic TSA of power systems, such as fault type, fault location, and FCT. Suitable distributions are used to model these factors. The modeling approaches are described as follows. Normally, shunt faults, such as three-phase (LLL), double-line-to-ground (LLG), line-to-line (LL) and single-line-to-ground (LG) short circuits, are considered for evaluating probabilistic transient stability. A discrete distribution is normally used to model the fault type. Based on past system statistics, a usual practice is to select the probability of LLL, LLG, LL and LG short circuits, as 0.05, 0.1, 0.15 and 0.7, respectively [33]. This paper uses the same approach. The probability distribution of fault location on a transmission line is usually assumed to be uniform, i.e., the fault occurs with equal probability at any point along any line of the test network [10]. This paper uses the same approach. The procedure of fault clearing constitutes of three stages: fault detection, relay operation and breaker operation. If the primary protection and breakers are fully reliable, the clearing time is the only uncertain entity. A Normal distribution is generally used to model this time [10]. In this paper, fault is applied at 1 s and it is cleared, after a mean time of 0.2 s (12 cycles for a 60 Hz system) and standard deviation of 0.005s (based on the Normal distribution). The detailed computation procedure is shown in Fig. 5. In the first step, load flow is run on the base case under normal condition, i.e., without any fault. Then, a random line is selected for a fault. The line is selected based on uniform distribution as described above. Then, fault type and fault location on the line are selected. Fault in then cleared based on FCT Normal distribution. The transient stability time domain simulation is done for 10 s (which is usually the time for large disturbance power system stability study). After making sure that there are enough samples for convergence, risk indices for angle, voltage, and frequency are computed. Eventually, global instability risk index is computed. In this work, it was found that 30,000 MC samples are enough for convergence. The value of mean was used as the convergence criteria. The values of $R_{AM}$, $R_{VM}$, $R_{FM}$, and $G$ against $N$ are shown in Table 1. From this Table, it is evident that 30,000 samples are sufficient to use in MC simulation. This is because further increasing the samples does not change the mean values: $R_{AM}$, $R_{VM}$, $R_{FM}$ and $G$. Thus, convergence is achieved.

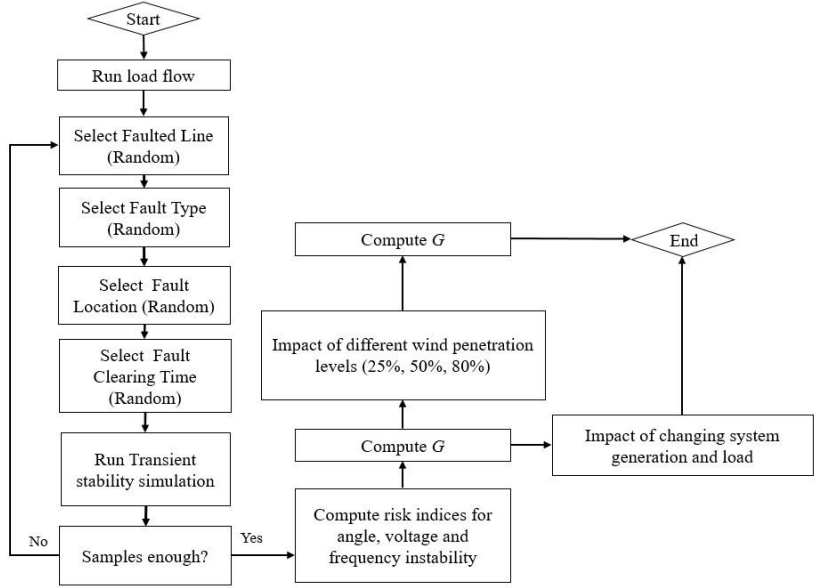

Figure 5. Procedure for computing *G*

Table 1. Values of *G* against *N*

| N | $R_{AM}$ | $R_{VM}$ | $R_{FM}$ | G |
|---|---|---|---|---|
| 1,000 | 0.33 | 0.25 | 0.25 | 0.33 |
| 5,000 | 0.35 | 0.26 | 0.27 | 0.35 |
| **30,000** | **0.36** | **0.27** | **0.29** | **0.36** |
| 50,000 | 0.36 | 0.27 | 0.29 | 0.36 |

## 5. Modeling of Generation Sources

*A. Synchronous Generator*

Standard 6$^{th}$ order model is used for modeling all synchronous generators. All synchronous machines include (*TGOV1*) turbine governor, (*IEEEX1*) exciter and (*STAB1*) power system stabilizer. The associated mathematical model for a standard 6$^{th}$ order synchronous generator is given by following equations [34].

$$T_J \frac{d\omega}{dt} = M_m - M_c - D(\omega - \omega_o) \qquad (17)$$

$$\frac{d\delta}{dt} = \omega - \omega_o \tag{18}$$

$$T'_{do}\frac{dE'_q}{dt} = E_{fd} - (x_d - x'_d)I_d - E'_q \tag{19}$$

$$T''_{do}\frac{dE''_q}{dt} = -E''_q - (x'_d - x''_d)I_d + E'_q + T''_{do}\frac{dE'_q}{dt} \tag{20}$$

$$T'_{qo}\frac{dE'_d}{dt} = -E'_d + (x_q - x'_q)I_q \tag{21}$$

$$T''_{qo}\frac{dE''_d}{dt} = -E''_d - (x'_q - x''_q)I_q + E'_d + T''_{qo}\frac{dE'_d}{dt} \tag{22}$$

Where $T_J$ is generator inertia constant; $M_m$ is mechanical torque; $M_e$ is electromagnetic torque; D is damping coefficient; $\omega$ is rotor speed; $\omega_o$ is synchronous speed; $E_{fd}$ is field voltage; $E'_d$ and $E'_q$ are d and q axis components of transient electric potential; $E''_d$ and $E''_q$ are d and q axis components of subtransient electric potential; $x_d, x'_d, x''_d, x_q, x'_q, x''_q$ are d and q axis synchronous reactance, transient reactance and subtransient reactance, respectively; $T'_{do}, T'_{qo}$ are d and q axis transient time constants, and $T''_{do}, T''_{qo}$ are d and q axis subtransient time constants.

## C. Doubly Fed Induction Generator (DFIG)

A reduced 3rd order model, which neglects the stator transients, is used to represent DFIGs. The model has a structure like that proposed by WECC [35] and IEC [36]. Ignoring the stator current dynamics, the mathematical equations governing the DFIG model are as follows [37].

$$\frac{1}{\omega_b}\frac{de_d}{dt} = \frac{-1}{T_o}[e_d - (X - X')i_{qs}] + s\omega_s e_q - \omega_s \frac{L_m}{L_m + L_r}v_{qr} \tag{23}$$

$$\frac{1}{\omega_b}\frac{de_q}{dt} = \frac{-1}{T_o}[e_q - (X - X')i_{ds}] - s\omega_s e_d + \omega_s \frac{L_m}{L_m + L_r}v_{dr} \tag{24}$$

$$\frac{d\omega_r}{dt} = \frac{1}{2H_g}[K_{tw}\theta_{tw} + D_{tw}(\omega_t - \omega_r) - (e_d i_{ds} + e_q i_{qs})] \tag{25}$$

$$v_{ds} = -r_s i_{ds} + X'i_{qs} + e_d \tag{26}$$

$$v_{qs} = -r_s i_{qs} - X' i_{ds} + e_q \tag{27}$$

$$P_w = v_{ds} i_{ds} + v_{qs} i_{qs} - v_{dr} i_{dr} - v_{qr} i_{qr} \tag{28}$$

$$Q_w = v_{qs} i_{ds} - v_{ds} i_{qs} \tag{29}$$

Where $e_d, e_q$ are d and q components of internal voltage; $P_w, Q_w$ are active and reactive power of DFIG absorbed by the network; $X, X'$ are open-circuit and short-circuit reactance; $T_o$ is the transient open-circuit time constant; $H_g$ is generator inertia constant; $\omega_b, \omega_s, \omega_r$ are system base speed, synchronous speed and rotor speed, respectively; $s$ is generator slip; $\theta_{tw}$ is the shaft twist angle (radians); $K_{tw}, D_{tw}$ are the shaft stiffness and mechanical damping coefficients; $v_{ds}, v_{qs}$ are stator voltages; $v_{dr}, v_{qr}$ are rotor voltages; $i_{ds}, i_{qs}$ are stator currents; $i_{dr}, i_{qr}$ are rotor currents; $r_s$ is stator resistance; $L_m, L_r$ are magnetizing and rotor inductances.

## 6. Case Studies and Simulations

The well-known IEEE 39-bus test transmission system, also known as the 10-machine New England test system, operating at 345 kV, was used to conduct the required simulations. The numerical data and parameters were taken from [38]. This system is a good choice for the study as it has been widely used by various researchers for studying large disturbance stability phenomenon in power transmission systems [39-41]. The system one-line diagram is shown in Fig. 6. The network consists of 10 synchronous generators, 34 transmission lines and 12 transformers. It has 19 constant impedance loads totaling 6097.1 MW. The installed capacity of the generators is 6140.81 MW. Transmission lines are modelled using the standard approach of equivalent pi circuit of a long transmission line. Every circuit breaker is represented by an ideal switch which can open at current zero crossings. All wind generators are assumed to be operating at their maximum outputs. This is a reasonable assumption as majority of stability issues can be investigated under this scenario, due to maximum acceleration of wind generators [42]. Thus, no variations in wind generation has been considered in this work. Initial operating conditions were obtained based on load flow analysis. Further, it is assumed the system load is at its peak value. Moreover, the period of transient stability (10 s) makes variation of load irrelevant as load changes are normally recorded at every 10 or 15-minute intervals. All time-domain (dynamic) simulations are RMS simulations which are performed with the help of DIgSILENT PowerFactory commercial software (version 2018) [43].

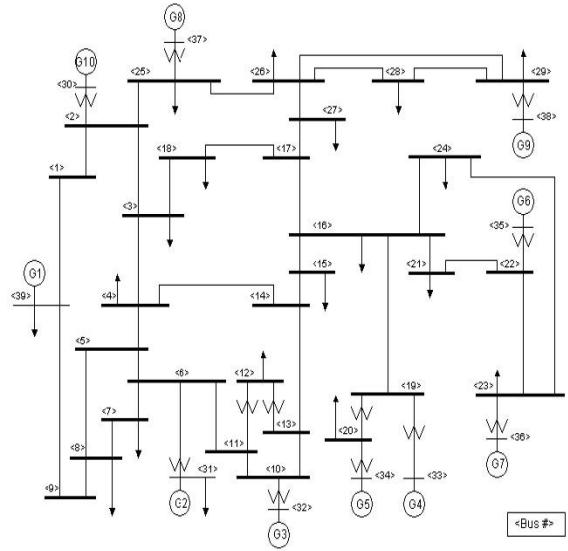

Figure 6. IEEE 39-bus test system

## 7. Results and Discussion

Based on the computation procedure described in Section 4, values of $R_{AM}$, $R_{VM}$, $R_{FM}$, and eventually $G$ were computed. In the next step, synchronous generation was replaced by wind generation and corresponding values of $G$ were computed. The penetration level in this paper is defined as follows.

$$\% \text{ Penetration level} = \frac{\text{Total renewable generation (MW)}}{\text{Total system generation (MW)}} \times 100\% \qquad (30)$$

To achieve the required penetration levels, selected synchronous generators (SGs) are replaced by renewable generation sources. Table 2 shows the generators which are replaced by renewable wind generations to achieve the required penetration level.

Table 2. Description of penetration levels

| DFIG Penetration level (%) | SGs replaced |
|---|---|
| 25 | G1, G3 |
| 50 | G1, G3, G5, G9, G10 |
| 80 | G1, G3, G4, G5, G6, G7, G9, G10 |

For instance, 25% DFIG penetration means that equivalent amount of synchronous generation (G1 and G3) in the system is replaced by DFIG (with the same MW and MVA rating). Similar procedure is used to achieve other penetration levels. Fig. 7 shows the value of *G* for each MC sample. From Fig. 7, a distribution can be derived which is shown in. This is a Normal (Gaussian) Distribution with a mean value of 0.36 and a standard deviation of 0.05. Similar distributions for *G* are obtained for other cases as shown in Figs. 8-10.

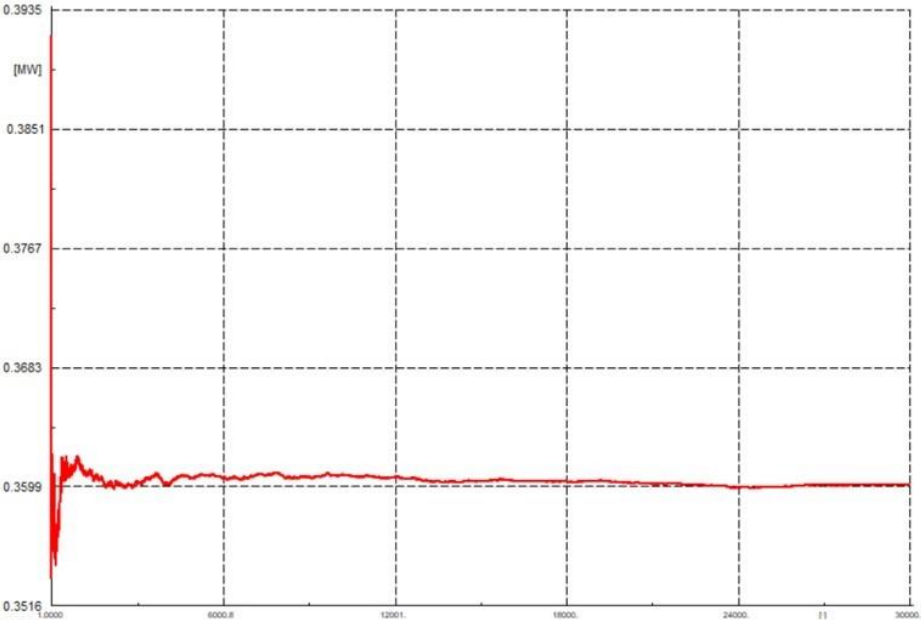

Figure 7. Variation of *G* with *N* (base case)

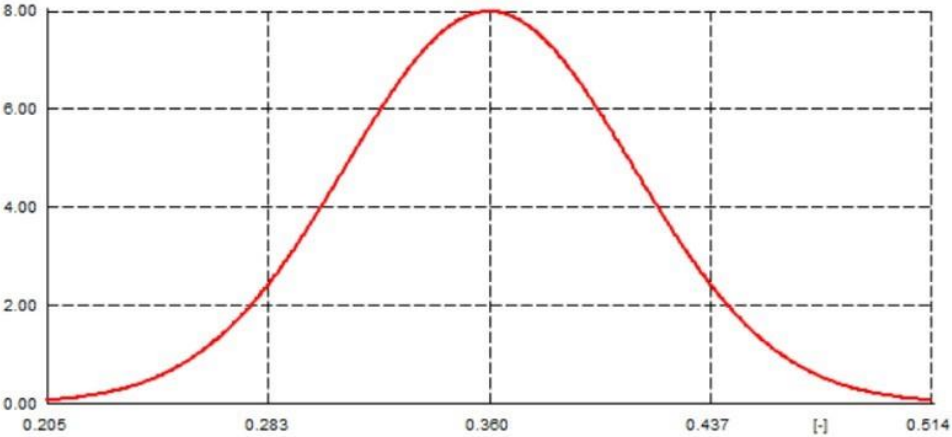

Figure 8. Normal distribution (mean 0.36) of *G* (base case)

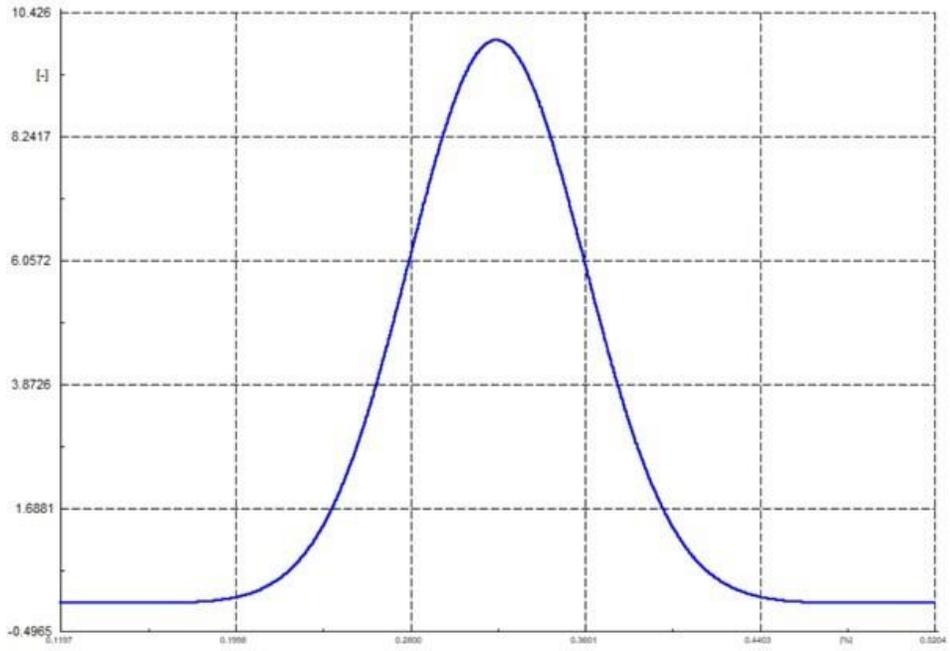

Figure 9. Normal distribution (mean 0.32) of *G* (25% wind)

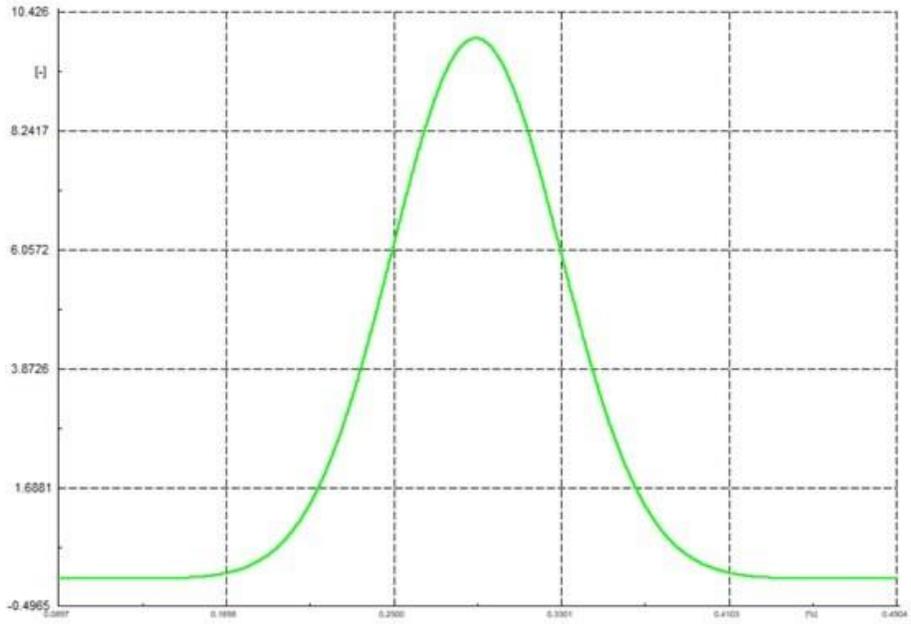

Figure 10. Normal distribution (mean 0.29) of *G* (50% wind)

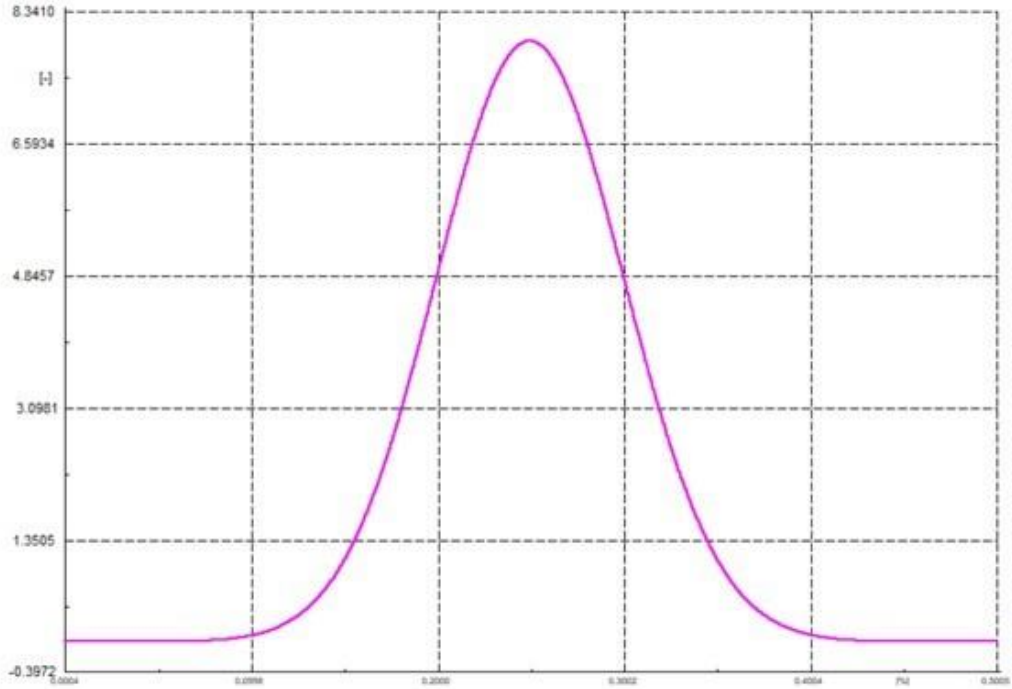

Figure 11. Normal distribution (mean 0.25) of *G* (80% wind)

From Figs. 8-11, value of *G* can easily be computed, i.e., the mean values of the corresponding Normal distribution. These values are shown in Table 3 and graphed in Fig. 12.

Table 3. Values of *G* for different cases

| Case type | *G* |
|---|---|
| Base case | 0.36 |
| 25% wind | 0.32 |
| 50% wind | 0.29 |
| 80% wind | 0.25 |

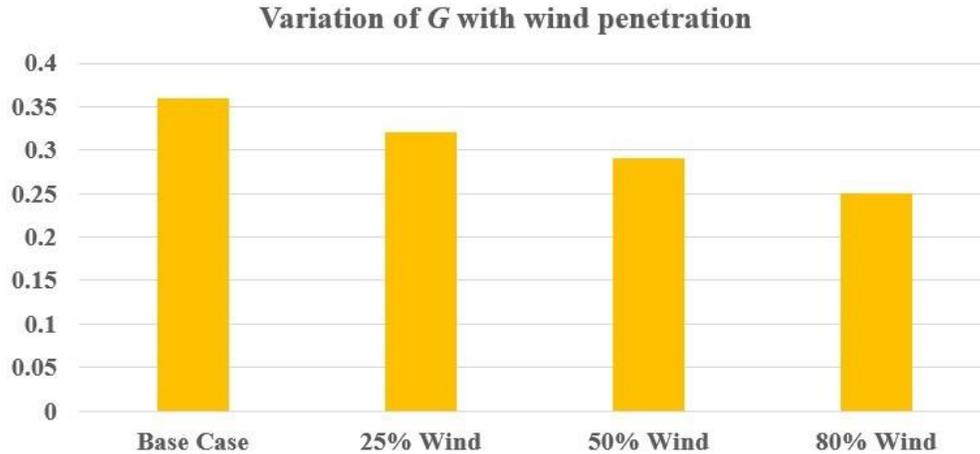

Figure 12. Variation of $G$ with wind penetration

It is evident from Fig. 12, that the addition of DFIG in conventional power system (having only synchronous generation) improves the overall system stability. This is because the value of $G$ decreases with increasing wind penetration. A possible reason is the superior reactive power control capabilities of DFIG during faults.

It is of great interest to observe the trend of increasing system generation and load on the value of $G$. Fig. 13 shows the impact of increasing system generation (keeping system load constant) on $G$. As system generation is increased, value of $G$ starts to decrease. Hence, system tends to be less risky with respect to stability if system generation is plentiful. However, adding generation is expensive and care must be taken before doing that. It must be made sure that cost-benefit analysis is conducted before adding new generation n the system. Fig. 14 shows the impact of increasing system load (keeping system generation constant) on $G$. As system load rises, there is a sharp hike in the value of $G$. This is understandable as there is not adequate generation to fulfil the system load demand. This makes the system riskier with respect to stability when there is a fault. Hence, power system operator must exercise caution to assure that critical loss of load does not occurs and if does occur, it must be restored as quickly as possible.

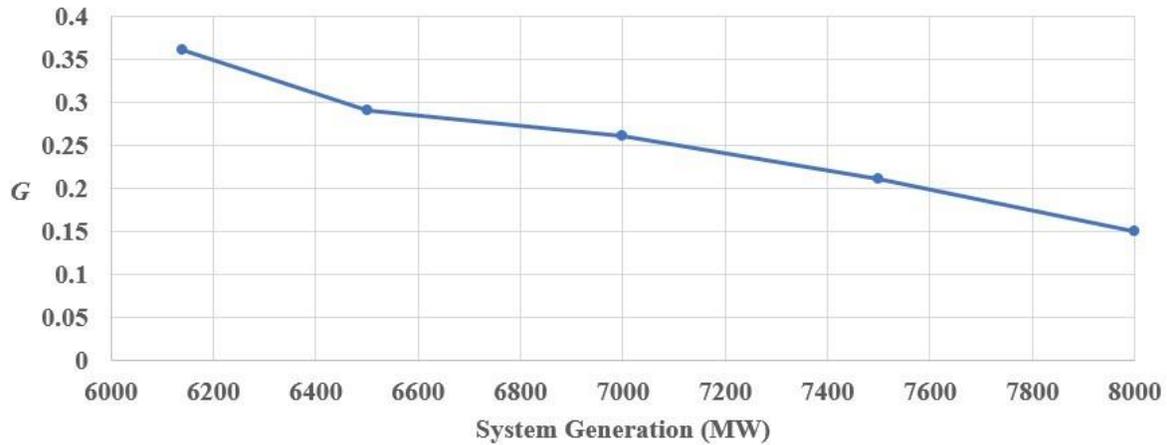

Figure 13. Variation of *G* with increasing system generation

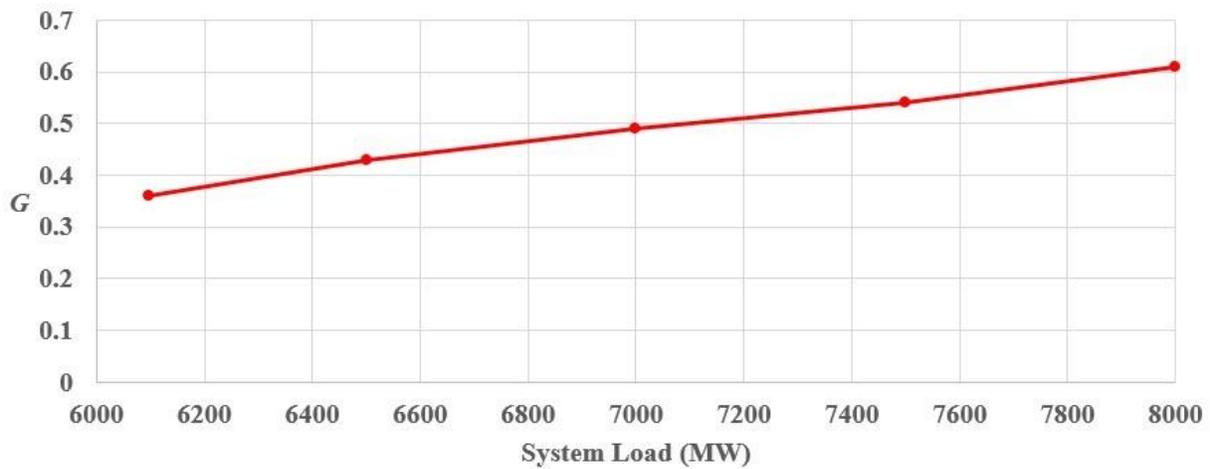

Figure 14. Variation of *G* with increasing system load

## 8. Conclusion and Future Work

The work proposed a global instability risk index for a power system in the presence of renewable wind generation. IEEE 39-bus test system was used to conduct required simulations. The results showed that adding DFIG wind penetration enhances the overall stability of power system under fault conditions. Also, it was determined that increasing system generation is beneficial for power system stability. On the contrary, decreasing system load is detrimental for power system stability. Moreover, the work highlighted the significance of using risk-based stability criterion in planning and operation procedures of a typical power system. As a future work, the impact of PV renewable generation can be observed on the proposed global instability index. Moreover, additional indices can be devised for quantifying the global instability index, incorporating both small and large disturbances. Multiple contingencies, including generator and transformer contingencies, can be considered to assess the overall risk-based system instability.